\documentclass[twocolumn,preprintnumbers,nofootinbib,prl]{revtex4}

\usepackage{amssymb,amsmath,graphicx}
\usepackage{epsf}
\usepackage{graphicx,epsfig}
\usepackage{amsfonts}
\usepackage{amssymb}
\usepackage[hypertex]{hyperref}

\def\bp{{\bf p}}

\def\bx{{\bf x}}

\def\CH{{\cal H}}

\def\CO{{\cal O}}

\def\ttau{{\tilde \tau}}

\def\half{\frac{1}{2}}

\begin{document}

\newcommand{\be}{\begin{equation}}
\newcommand{\ee}{\end{equation}}
\newcommand{\bea}{\begin{eqnarray}}
\newcommand{\eea}{\end{eqnarray}}
\newcommand{\barr}{\begin{array}}
\newcommand{\earr}{\end{array}}

\pagestyle{plain}


\title{Large Non-Gaussianities with Intermediate Shapes from
  Quasi-Single Field Inflation}

\author{Xingang Chen$^{1,2}$ and Yi Wang$^{3,4}$}

\affiliation{
$^1$Center for Theoretical Physics, Massachusetts
  Institute of Technology, Cambridge, MA 02139, USA\\
  $^2$Center for Theoretical Cosmology, DAMTP, University of Cambridge, Cambridge CB3 0WA, UK\\
$^3$Physics Department, McGill University, Montreal, H3A2T8, Canada
\\
$^4$KITPC, Institute of Theoretical Physics, Chinese Academy of Sciences, Beijing 100190, China
}


\begin{abstract}

We study the slow-roll inflation models, where the inflaton slow-rolls
along a trajectory whose orthogonal directions are lifted by
potentials with masses of order the Hubble parameter. In these
models large
non-Gaussianities can be generated through the transformation from the
isocurvature modes to the curvature mode, once the inflaton
trajectory turns. We find large bispectra with a one-parameter family of
novel shapes, interpolating between the
equilateral and local shape.
According to the in-in formalism, the shapes of these
non-Gaussianities
are different from a simple projection from the isocurvature
non-Gaussian correlation functions.

\end{abstract}
\maketitle

\section{Introduction}

Inflation \cite{Guth:1980zm} is the leading candidate for creating the
homogeneous and
isotropic universe and generating the primordial density fluctuations
for large scale structure. The condition for
inflation to happen is
that the inflatons stay near the top of the potential for a
sufficiently
long time, so that the vacuum energy drives the accelerated expansion of
the Universe.

Generically, it is found that such a condition
needs to be fined-tuned, at least for various types of models that
have
sensible UV completion \cite{Copeland:1994vg}. For slow-roll
inflation, this means that the
generic inflaton potential has a steep shape, characterized by a mass
of order the Hubble parameter $H$.
On the other hand, it is general that
the inflaton can roll in a space with multiple light fields.
So at least one of the
directions needs to be fine-tuned. A combination of these two
aspects suggests a generic situation -- the inflaton slow-rolls
along a trajectory whose orthogonal directions are lifted
by potentials with masses of order $H$. We call this class of models
the quasi-single field inflation. If the inflaton trajectory is
straight,
this is equivalent to the single field inflation, which generates
unobservably small
primordial non-Gaussianities \cite{Maldacena:2002vr}. However, as we
will show
in this paper, once the trajectory turns, large non-Gaussianities
with novel shapes
can be generated, which are potentially detectable in current and
future experiments.

We call the mode in the tangential direction the curvature mode, and
the modes in the
perpendicular directions the isocurvature modes. 
The non-Gaussianities originated from the curvature mode are very small because the potential is flat.
Since the
isocurvature modes in our model are massive, $m \sim H$, and
can have large higher order interactions,
the non-Gaussian correlation functions of the
isocurvature
modes can be very large. However, in dS space the amplitude of the
quantum fluctuations of a
massive field decays exponentially after horizon exit,
and is also fast oscillating if $m \gg H$.
This is why such modes are usually
not considered. But for $m\sim H$, the suppression due to oscillation
is only
marginal. In the mean while, although the isocurvature amplitude still
decays, the part of it that is transferred to the
tangential direction through turning trajectory becomes part of the
curvature mode, which remains constant after horizon exit. The
non-Gaussianities in the isocurvature modes are thus transferred in
this way. This is the
main physical picture behind our calculation.

We use the in-in formalism \cite{Weinberg:2005vy}
to compute the precise momentum dependence
(shapes) of the bispectra. In this formalism,
the natural way to implement the transformation from the isocurvature
to curvature mode is to introduce a two-point transfer vertex
between the two modes. It is then straightforward to do a perturbative
calculation for correlation functions
according to the usual Feynman diagrams.
As we will see, the effect of the transfer vertex is not a simple
projection of the
three-point function of the isocurvature modes.

During the computation, we find that two equivalent representations of
the in-in formalism, namely
the original factorized form and the commutator
form, are computationally advantageous in complementary ways.
In certain parameter space, each representation
encounters spurious divergence either in IR or UV.
These divergences are
called spurious because they will eventually be cancelled, but
significantly complicate the analytical and numerical
calculations. They are completely absent, however, when viewed in a
different representation.

We find a one-parameter family of bispectrum shapes, lying
in-between the well-known equilateral and local shape. The shape
sensitively depends on the mass of the isocurvature mode; the
size depends on the turning angular velocity
and the cubic interaction strength among isocurvature modes.

The mechanism that the isocurvature modes, in particular their non-Gaussianities, can be transferred to the curvature mode through turning trajectory has been pointed out and studied intensively in the past \cite{Salopek:1990jq,Gordon:2000hv,Bartolo:2001vw,Bernardeau:2002jy,Sasaki:1995aw}.
The focus was on the light isocurvature modes with mass much less than $H$. These modes do not decay after horizon-exit and non-Gaussianities come from non-linear classical evolution of super-horizon modes in multifield space. The resulting shapes of non-Gaussianities are local.
Also note that for multifield slow-roll inflation models, it is found to be very difficult to generate large non-Gaussianities through this mechanism because the restrictive slow-roll conditions in all directions.
Here in the quasi-single field inflation models, we focus on massive isocurvature fields, which, as we emphasized and will show in details later, are crucial for generating large non-Gaussianities with new intermediate shapes.

\section{The model, mode functions and transfer vertex}

We consider a two-field model as an example. It is convenient to
write the action in terms of the fields in polar coordinates, $\theta$ and $\sigma$,
that are tangential and orthogonal to the turning trajectory,
respectively,
\bea
S = \int d^4x \sqrt{-g} \left[
\half (R+\sigma)^2 g^{\mu\nu} \partial_\mu \theta \partial_\nu \theta
+ \half g^{\mu \nu} \partial_\mu \sigma \partial_\nu \sigma \right.
\cr \left.
- V_{sr}(\theta) - V(\sigma) \right] ~,~
\eea
where $R$ is the radius of the turning trajectory. The
$V_{sr}(\theta)$ is a usual slow-roll potential and we choose the
rolling velocity $\dot
\theta_0 >0$. The potential
\bea
V(\sigma) = \half m_\sigma^2 \sigma^2 + \frac{1}{6} V''' \sigma^3 +
\cdots
\eea
traps the $\sigma$ field at $\sigma =0$
\cite{ft_sigma}. In principle, the parameters that characterize the
turning trajectory, $\dot \theta_0$ and
$R$, and the cubic interaction strength $V'''$, can vary along the
trajectory. In
this paper we consider the constant turn case in which they are all
constant.

We perform the perturbation in
the gauge where the scale factor $a(t)$ is homogeneous.
The leading kinematic Hamiltonian density for the quantum
fluctuations,
$\delta\theta_I$ and $\delta\sigma_I$, in
the interaction picture is
\bea
\CH_0 &=& a^3 \left[ \half R^2 \dot {\delta\theta_I}^2 +
  \frac{R^2}{2a^2}
  (\partial_i \delta\theta_I)^2
\right.
\cr
&+&
\left.
\half \dot{\delta\sigma_I}^2 + \frac{1}{2a^2} (\partial_i
\delta\sigma_I)^2
+ \half (m_\sigma^2 + 7 \dot\theta_0^2) \delta \sigma_I^2
\right] ~. ~
\eea
The leading interaction Hamiltonian density is
\bea
\CH^I_2 &=& -c_2 a^3 \delta\sigma_I \dot{\delta\theta_I} ~,
\label{CH2}
\\
\CH^I_3 &=&  c_3 a^3 \delta\sigma_I^3 ~,
\label{CH3}
\eea
where
$c_2 = 2 R \dot\theta_0$, $c_3= V'''/6$
are constants.

We quantize the fields $\delta \theta_I$ and $\delta
\sigma_I$ in the momentum space $\bp$,
$\delta\theta_\bp^I = u_p a_\bp + u_{-p}^* a_{-\bp}^\dagger$,
$\delta\sigma_\bp^I = v_p b_\bp + v_{-p}^* b_{-\bp}^\dagger$,
where $a_\bp$ and $b_\bp$ are independent of each other and each
satisfies the usual commutation relation.
The mode functions in terms of the conformal
time $\tau= \int dt/a(t)$ are
\bea
u_p &=& \frac{H}{R\sqrt{2p^3}} ( 1+i p\tau)e^{-i p\tau} ~,
\label{mode_u}
\\
v_p &=& -i e^{i(\nu+\half)\frac{\pi}{2}} \frac{\sqrt{\pi}}{2} H
  (-\tau)^{3/2} H^{(1)}_\nu (-p\tau) ~,
\label{mode_v1}
\eea
where $\nu = \sqrt{9/4-m^2/H^2}$ and
$m^2=m_\sigma^2+7\dot\theta_0^2$.
As mentioned in the Introduction,
the amplitude for the isocurvature mode $v_p$ decays as
$(-\tau)^{3/2-\nu}$ at late time $\tau\to 0$.
If $m^2/H^2>9/4$, $\nu$ is imaginary,
$v_p$ has an additional fast oscillating factor,
$\sim e^{\nu \ln(-p\tau/2)}$,
even after the horizon exit. This suppresses its contribution.
We will consider the case $0 \le \nu<3/2$, i.e.~$9/4 \ge m^2/H^2>0$,
in this paper.

\begin{figure}[t]
\begin{center}
\epsfig{file=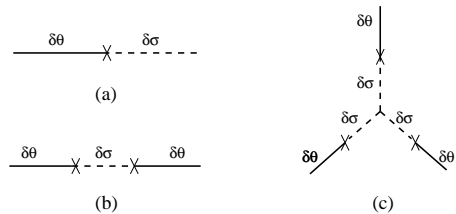, width=6cm}
\end{center}
\medskip
\caption{Feynman diagrams for the transfer vertex (a), and the
  corrections from the isocurvature mode
  to the
  power spectrum (b) and bispectrum (c).}
\label{Fdiagrams}
\end{figure}

The term that is responsible for the transformation from the
isocurvature to curvature mode is (\ref{CH2}). This introduces
the ``transfer vertex'' (Fig.~\ref{Fdiagrams} (a)).
The contribution from isocurvature to curvature
correlation functions can therefore be computed
according to the Feynman diagrams in Fig.~\ref{Fdiagrams}
(b)(c).
The term (\ref{CH3}) is the leading source for large
non-Gaussianities, since all other cubic interactions in the expansion
involve $\delta \theta$ which is in the slow-roll direction.

\section{Bispectra}

We compute the three-point correlation function $\langle \delta
\theta^3 \rangle$.
The bispectrum $\langle \zeta^3 \rangle$ is related to it through the
usual time-delay relation $\zeta \approx -H \delta\theta/\dot\theta$.

The in-in formalism \cite{Weinberg:2005vy} gives
\bea
\langle \delta\theta^3 \rangle &\equiv&
\langle 0| \left[ \bar T \exp\left( i\int_{t_0}^t d\tilde t'
  H_I(\tilde t')\right)
  \right] \delta\theta_I^3(t)
\cr
&& ~~~~
  \left[ T \exp\left( -i\int_{t_0}^t dt' H_I(t')\right)
  \right] |0\rangle ~,
\label{inin}
\eea
where $t$ is the end of the inflation, and $t_0$ is the infinite
past.
The $H_I$ is the interaction Hamiltonian,
\bea
H_I = \int d^3 \bx ~ (\CH^I_2+\CH^I_3)
~.~
\eea
One can directly expand
the exponentials in (\ref{inin}). For our purpose, the relevant
terms are those that
involve four $H_I$'s. Each resulting quartic integral contains two
``factors''. Each factor has an (anti-)time-ordered integration but
there is no cross
time-ordering between the two. We call
this form the ``factorized form''. One can also rewrite all the
integrands into one single time-ordered quartic integral,
\bea
&&\int_{t_0}^t dt_1 \int_{t_0}^{t_1} dt_2 \int_{t_0}^{t_2} dt_3
\int_{t_0}^{t_3} dt_4
\cr
&&\langle [ H_I(t_4), [ H_I(t_3), [ H_I(t_2), [ H_I(t_1),
	\delta \theta_I(t)^3]]]] \rangle ~.~
\label{ComForm}
\eea
We call this the ``commutator form''.
The Feynman diagram Fig.~\ref{Fdiagrams} (c) corresponds to
replacing one of four $H_I$'s in (\ref{inin}) or (\ref{ComForm})
with $H_3^I$ and the rest three with $H_2^I$.
A straightforward exercise of Wick contraction leads to, for the
factorized form,
\bea
&-& 12 c_2^3 c_3 u_{p_1}^* u_{p_2} u_{p_3}(0)
\cr &\times&
{\rm Re} \left[ \int_{-\infty}^0 d\ttau_1~ a^3 v_{p_1}^*
u'_{p_1}(\ttau_1)
\int_{-\infty}^{\ttau_1} d\ttau_2~ a^4 v_{p_1}
v_{p_2} v_{p_3}(\ttau_2) \right.
\cr &\times& \left.
\int_{-\infty}^0 d\tau_1~ a^3 v_{p_2}^*
u_{p_2}^{\prime *} (\tau_1)
\int_{-\infty}^{\tau_1} d\tau_2~ a^3 v_{p_3}^*
u_{p_3}^{\prime *} (\tau_2) \right]
\cr &\times&
(2\pi)^3 \delta^3(\sum_i \bp_i) + {\rm 9~other~similar~terms}
\cr
&+& {\rm 5~permutations~of~} \bp_i ~;
\label{FacForm_term}
\eea
and for the commutator form,
\bea
&& 12 c_2^3 c_3 u_{p_1}(0) u_{p_2}(0) u_{p_3}(0)
\cr
&\times&
{\rm Re} \left[ \int_{-\infty}^0 d\tau_1 \int_{-\infty}^{\tau_1}
  d\tau_2 \int_{-\infty}^{\tau_2} d\tau_3 \int_{-\infty}^{\tau_3}
  d\tau_4~
\prod_{i=1}^4 \left( a^3(\tau_i) \right)
\right.
\cr
&\times&
a(\tau_2)
\left(u_{p_1}^{\prime}(\tau_1) - c.c. \right)
\left( v_{p_1}(\tau_1) v_{p_1}^*(\tau_2) - c.c. \right)
\cr
&\times&
\left.
\left( v_{p_3}(\tau_2) v_{p_3}^*(\tau_4) u_{p_3}^{\prime *}(\tau_4) -
c.c. \right)
v_{p_2}(\tau_2) v_{p_2}^*(\tau_3) u_{p_2}^{\prime *}(\tau_3)
\right]
\cr
&\times&
(2\pi)^3 \delta^3(\sum_i \bp_i)
+  {\rm 2~other~similar~terms}
\cr
&+& {\rm 5~permutations~of~} \bp_i ~.
\label{ComForm_term}
\eea
We write the argument $\tau_i$ only once
if they are all the same in one integrand, and the prime denotes the
derivative respective to the conformal time $\tau_i$.
The full details are
presented in Ref.~\cite{ChenWang}.

It is subtle to evaluate these integrals.
Let us first look at the factorized form. In the UV limit, $\tau_i\to
-\infty$, the integrands are fast oscillating. For the Bunch-Davies vacuum, the convergence of the
integration is achieved by slightly
tilting the contour
clockwise or counter-clockwise into the imaginary plane, $\tau_i \to
-\infty(1\pm i\epsilon)$. In fact, if the integration is convergent at IR,
$\tau_i\to 0$, it is
numerically much more convenient to do a Wick rotation $\tau_i \to
i z_i$ for the left factor and opposite for the right. The integration
range for $z_i$ is
from $-\infty$ to $0$. Now it is explicit that
the integrand of each factor decays exponentially at UV.
This is the case for $0\le \nu<1/2$.

However, for $1/2<\nu<3/2$,
using the asymptotic behavior of the mode functions, it is easy to see
that each term in (\ref{FacForm_term}) is IR divergent.
Physically, larger $\nu$ corresponds to smaller $m$ for the
isocurvature mode. So the mode decays slower.
The conversion from isocurvature to curvature
mode lasts longer
after the horizon exit. As we will see, this causes slower convergence
in the IR, but not divergence.

Let us look at this in the commutator form (\ref{ComForm_term}).
Due to the subtraction of
the complex conjugate in various terms, the integrand decreases faster
in IR.
For example, $u'_{p_1}(\tau_1)$ behaves as $(-\tau_1)$ as $\tau_1\to
0$; but after subtracting off its complex conjugate, we have
$(-\tau_1)^2$.
The next two terms are slightly more complicated but
similar. Since all these three terms are imaginary,
$v_{p_2}v_{p_2}^* u_{p_2}^{\prime *}$ in the 4th line
must be imaginary to make the whole integration
real. For $\tau_{2,3}\to 0$, the leading term of $v_{p_2}v_{p_2}^*
u_{p_2}^{\prime *}$ is real, so we need the subleading term in this
limit. This also
increases the IR convergence. Overall, it is not difficult to see that
the IR convergence is achieved for all $0\le \nu <3/2$. However,
the Wick rotation no
longer works in this form. The original integrands from the left and
right factor
have been multiplied together, so after Wick rotation, the
exponential decay of some factors are cancelled by the
exponential growth of the others.

To summarize, for $1/2< \nu <3/2$, the factorized form is well
behaved in UV but encounters spurious divergence in IR in the
intermediate step, while the
commutator form is well behaved in IR but the UV convergence becomes
tricky \cite{UV}. Therefore we have proved that the whole
integral has no divergence. In fact, to see both types of convergence
at once,
we can choose a cutoff $\tau_c$ in the middle, and make the
whole expression take the factorized form in the UV, $\tau_i <
\tau_c$, and commutator form in the IR, $\tau_i > \tau_c$
\cite{ChenWang}.
This provides an efficient way to numerically calculate the shapes of
the bispectra.

\section{Squeezed limit and shape ansatz}

\begin{figure}[h]
\begin{center}
\epsfig{file=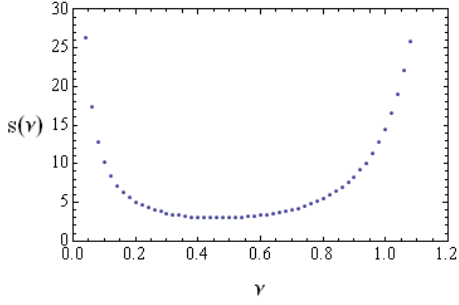, width=6cm}
\end{center}
\caption{The numerical coefficient
  $s(\nu)$ in the squeezed limit.}
\label{Fig_snu}
\end{figure}

We now look at the squeezed limit, $p_3\ll p_1=p_2$. In this limit,
simple analytical expressions for shapes are possible. This can
also help us construct simple analytical ansatz for the full shape,
to facilitate future data analyses.
In this limit, both
Eq.~(\ref{FacForm_term}) and (\ref{ComForm_term}) become
\bea
\frac{c_2^3 c_3}{HR^6}
\frac{1}{p_1^{\frac{7}{2}-\nu} p_2 p_3^{\frac{3}{2}+\nu}}
s(\nu) ~,
\label{slimit1}
\eea
where $s(\nu)$ is a $p_i$-independent
numerical number, but involve complicated integrals
\cite{ChenWang}. The result is presented in Fig.~\ref{Fig_snu}.

Some comments are in order, to explain the behavior of $s(\nu)$ near
$\nu=0$ and $\nu=3/2$. We have approximated the asymptotic behavior of
$H^{(1)}_\nu(-p_3\tau_i)$ in the small $p_3$ limit
as $-i (2^\nu \Gamma(\nu)/\pi)(-p_3 \tau_i)^{-\nu}$. For very small
$\nu\sim 0$, this requires $p_3/p_1 \ll e^{-1/\nu}$. So $p_3$ needs to
be
increasingly small as $\nu\to 0$. Otherwise, if we fix a small $p_3$,
near $\nu=0$
we should instead use $i (2/\pi) \ln (-p_3 \tau_i)$ as a better
approximation. Therefore the rising behavior in Fig.~\ref{Fig_snu}
near $\nu=0$
does not mean that the non-Gaussianities are blowing-up, rather
signals the change of the shape to
\bea
\sim \frac{\ln (p_3/p_1)}{p_1^{7/2} p_2 p_3^{3/2}}
~.
\label{slimit2}
\eea
As $\nu\to 3/2$, $m\to 0$,
the curvaton fluctuations do not decay, so its conversion to the curvature mode diverges in the constant turn case and an e-fold cutoff is necessary. Interestingly, in this limit, our shape approaches the local form, as expected from the multi-field models studied in Ref.~\cite{Salopek:1990jq,Gordon:2000hv,Bartolo:2001vw,Bernardeau:2002jy,Sasaki:1995aw}.
But here the non-Gaussianities can be made very large by having a large $V'''$, since we are not restricted to the slow-roll conditions even in the massless isocurvaton limit.

Combining (\ref{slimit1}) and (\ref{slimit2}), a good ansatz for the
full shape can be taken as, up to an overall amplitude,
\bea
- \frac{(p_1p_2p_3)^{-3/2}}{(p_1+p_2+p_3)^{3/2}}
N_\nu
\left( \frac{8 p_1 p_2 p_3}{(p_1+p_2+p_3)^3} \right) ~,~
\eea
where $N_\nu$ is the Neumann Function.
This ansatz gives an overall good match with the numerical results \cite{ChenWang}.

\begin{figure}[h]
\begin{center}
\epsfig{file=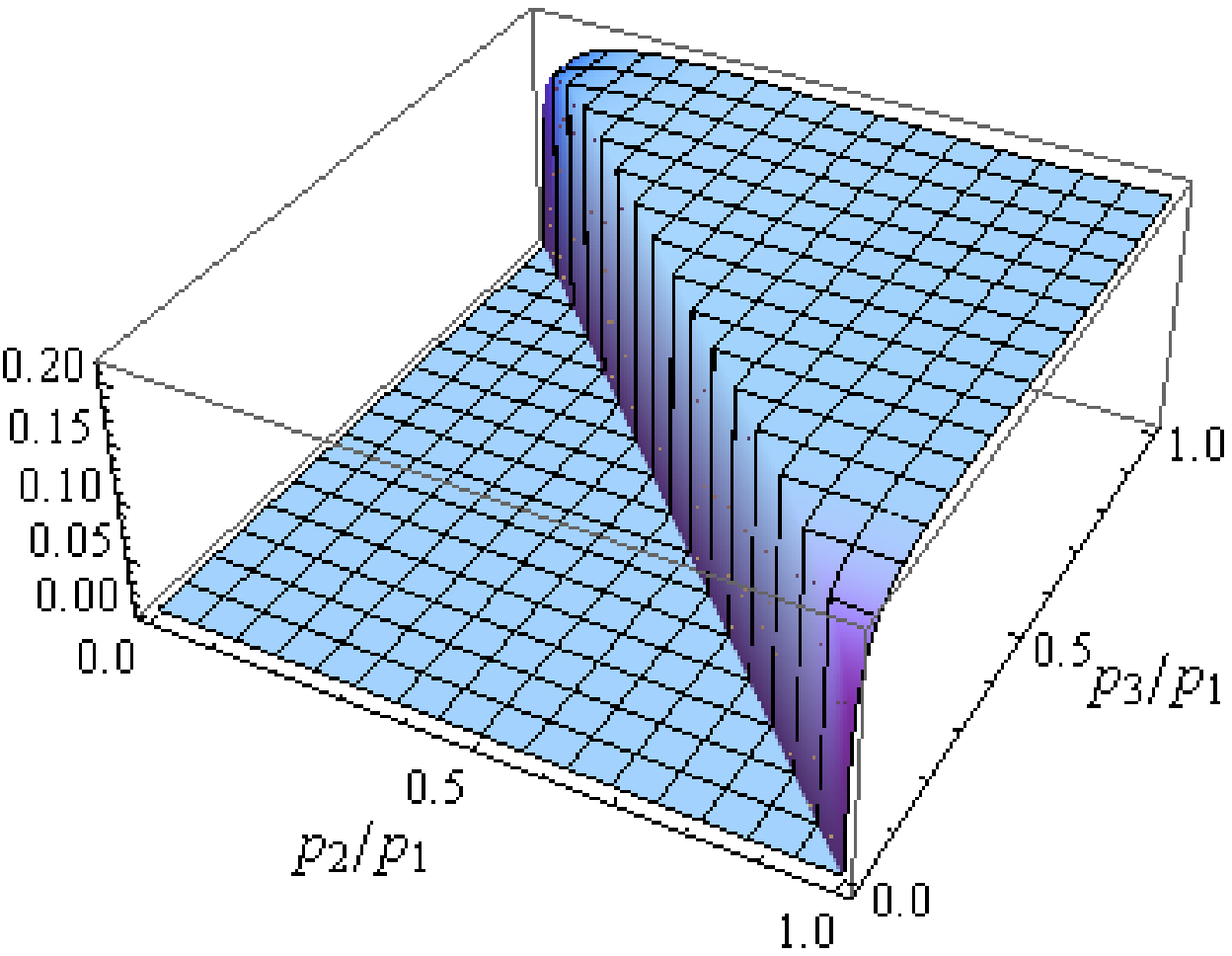, width=4.2cm}
\epsfig{file=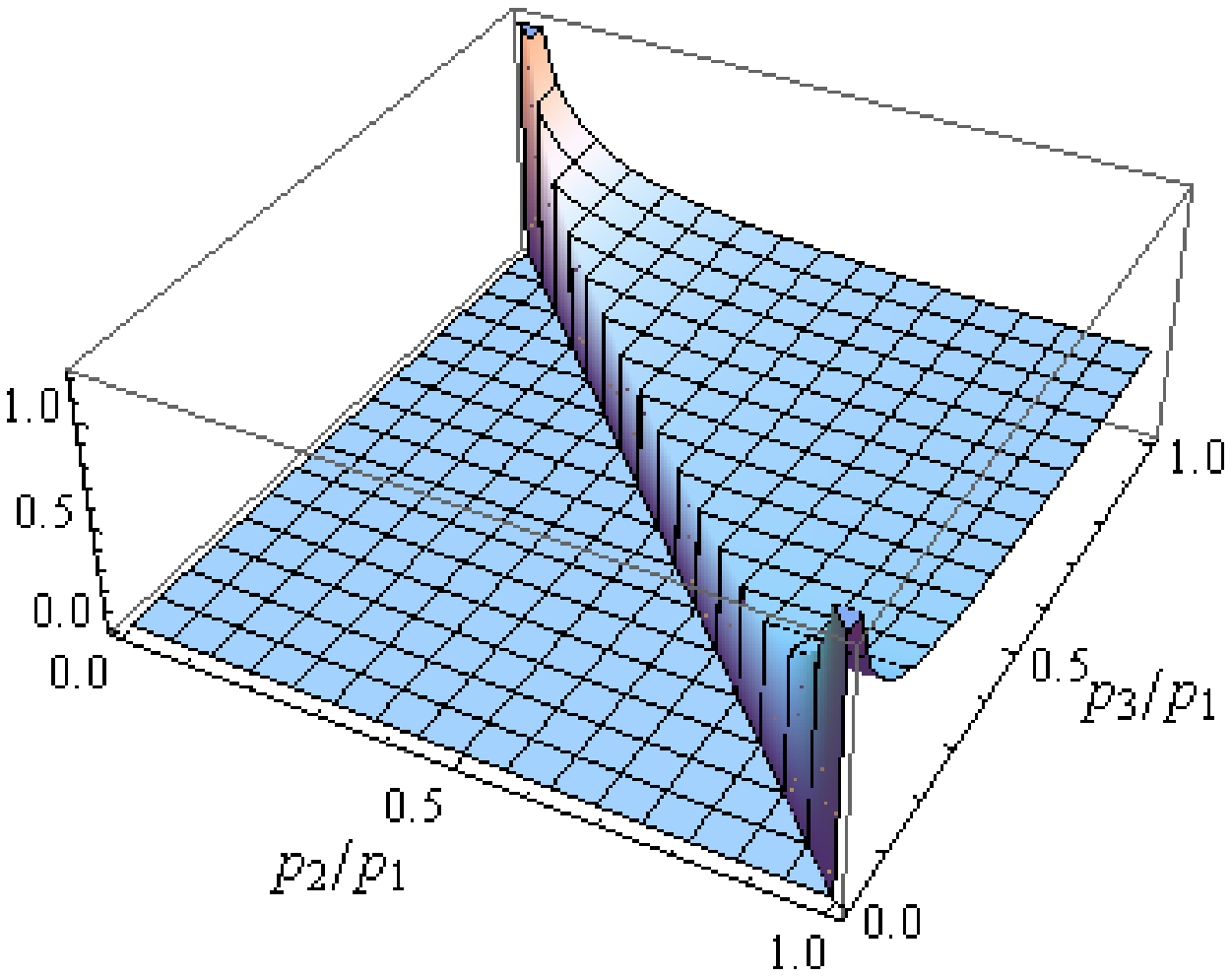, width=4.2cm}
\end{center}
\caption{Shapes of bispectra with intermediate form: 1)
  quasi-equilateral ($\nu=0.2$),
  2) quasi-local ($\nu=1$).
  The amplitudes are normalized by a factor of
  $(p_1p_2p_3)^2$ to be dimensionless.}
\label{shapes}
\end{figure}

Note that in the squeezed limit, this one-parameter family of shapes
goes as $p_3^{-3/2-\nu}$. This interpolates between the equilateral
form, $p_3^{-1}$, and the local form, $p_3^{-3}$ \cite{Babich:2004gb},
so we call it the ``intermediate form''.
Two examples of the shape ansatz are shown in
Fig.~\ref{shapes}.

For comparison, we look at the three-point function of the
isocurvature modes, $\langle \delta\sigma^3 \rangle$. Evaluating it
after the horizon exit, we find that its amplitude is decaying and its
shape goes as
$p_3^{-2\nu}$ in the squeezed limit. So at least in this model,
the shape of the correlation function is changed during the
transfer.
It is important to study this aspect in other multifield
models, such as \cite{Langlois:2008qf}.

\section{Size of non-Gaussianities}

The size $f_{NL}$ of a bispectrum is defined by taking the equilateral
limit \cite{Komatsu:2008hk},
\bea
\langle \zeta^3 \rangle \to \frac{9}{10} f_{NL} \frac{1}{p_1^6}
P_\zeta^2 (2\pi)^7 \delta^3(\sum \bp_i) ~,
\eea
where $P_\zeta$ is the power spectrum.
Using the relation $\zeta=-H \delta \theta/\dot \theta$ and $P_\zeta=
H^4/(4\pi^2 R^2 \dot \theta_0^2)\approx 6.1\times10^{-9}$,
we get
\bea
f_{NL}^{\rm int} = \alpha(\nu) \frac{1}{P_\zeta^{1/2}}
\left( \frac{-V'''}{H} \right) \left( \frac{\dot \theta_0}{H}
\right)^3
~.
\label{fNLint}
\eea
We investigate the order of magnitude of each factor.
The $\alpha(\nu)$ should be evaluated numerically, similar to
$s(\nu)$, but now in the equilateral limit. For example it is $\CO(1)$
and positive
near $\nu=0$. If we require that, in $V(\sigma)$, the quadratic term
dominates over the cubic interaction for $\sigma \lesssim H$, so that
we can trust the mode function, we need $|V'''|/H < (m_\sigma/H)^2
\sim
\CO(1)$. The perturbative method we used gives restriction on the size
of $\dot \theta_0/H$ because this parameter determines the strength of
the transfer vertex. For example, the correction to the power spectrum
can be simply calculated using Fig.~\ref{Fdiagrams} (b),
$\delta P_\zeta \sim (\dot
\theta_0/H)^2 P_\zeta$ and it is scale-invariant, so for it to be perturbative we need $(\dot
\theta_0/H)^2 \ll 1$.
It is possible that the non-Gaussianity is
larger if
$\dot \theta_0/H \sim 1$, but to trust the perturbative results in
this paper,
$\dot\theta_0/H<\CO(1)$.  Overall we see that $|f_{NL}^{\rm int}| \ll
\CO(10^4)$, and its sign is the opposite of $V'''$.
It will be very interesting to constrain it using the current and
future data \cite{Komatsu:2009kd}.
It is also interesting to study what the natural values
for $V'''$ and $\dot\theta_0$ are from
a more fundamental theory.

We thank B. Chen, A. Guth, Q-G. Huang, and M. Li for helpful
discussions and A. Frey for his
participation in the early stage of this work. XC was supported
by the US DOE under cooperative research
agreement DEFG02-05ER41360.
YW was supported by NSFC, NSERC and an IPP postdoctoral fellowship.


\begin{thebibliography}{99}
\bibitem{ft_sigma}
This requires $m_\sigma>\dot \theta_0$.
The $\dot \theta_0$ will cause a displacement for the vev, we can
redefine the $\sigma$ and $R$ so that
$\sigma=0$ is the effective minimum.
\bibitem{UV}
More detailed discussions on the computational advantages and disadvantages of both forms can be found in Ref.~\cite{ChenWang}.
\end{thebibliography}
\end{document}